\documentclass[aps,twocolumn,showpacs]{revtex4}
\usepackage{graphicx}
\def\bsigma{\mbox{\boldmath $\sigma$}}

\def\bpi{\mbox{\boldmath $\pi$}}
\def\sbpi{\mbox{\boldmath ${\scriptstyle \pi}$}}
\def\bOmega{\mbox{\boldmath $\Omega$}}

\begin{document}
\title{Spin magnetotransport in two-dimensional hole systems}
\author{O. E. Raichev}
\affiliation{Institute of Semiconductor Physics,
National Academy of Sciences of Ukraine,
Prospekt Nauki 45, 03028, Kiev, Ukraine}
%\date{\today}

\begin{abstract}
Spin current of two-dimensional holes occupying the ground-state subband
in an asymmetric quantum well and interacting with static disorder potential
is calculated in the presence of a weak magnetic field ${\bf H}$ perpendicular
to the well plane. Both spin-orbit coupling and Zeeman coupling are taken into
account. It is shown that the applied electric field excites both the transverse
(spin-Hall) and diagonal spin currents, the latter changes its sign at a finite
$H$ and becomes greater than the spin-Hall current as $H$ increases. The effective
spin-Hall conductivity introduced to describe the spin response in Hall bars is
considerably enhanced by the magnetic field in the case of weak disorder and
demonstrates a non-monotonic dependence on $H$.
\end{abstract}


\maketitle

One of the most challenging problems in the physics of two-dimensional (2D)
electron systems is the excitation of spin currents by an electric field
${\bf E}$ directed in the 2D plane. This phenomenon exists owing to the spin-orbit (SO)
interaction, which brings spin-dependent terms to the Hamiltonian of free electrons
(intrinsic SO coupling) as well as spin-dependent corrections to the scattering potential
(extrinsic SO coupling); see Ref. 1 for a review. In the quantum wells
grown along [001] crystallographic direction in cubic crystals of zinc-blende
type, the symmetry allows both non-diagonal (perpendicular to ${\bf E}$) and
diagonal (parallel to ${\bf E}$) currents of $z$-polarized spins. The diagonal
spin current does not appear if the SO splitting is isotropic.
The presence of non-diagonal spin current leads to the spin Hall
effect, which has been detected both in electron$^{2,3}$
and hole$^{4,5}$ systems by observing spin accumulation near the sample
boundaries. The original theoretical proposal$^{6}$ of the intrinsic spin Hall
effect has been based on the Rashba Hamiltonian$^{7}$ describing the SO
coupling in electron systems due to structural inversion asymmetry. However,
theoretical calculations$^{8}$ have proved the absence of static intrinsic
spin currents for this case, and this statement remains true when a magnetic
field is applied to the system.$^{9}$ Consideration of the equation of motion
for the spin density operator$^{10}$ allows one to extend applicability of this
result of Refs. 8 and 9 to any electron system described by the ${\bf p}$-linear
SO coupling. On the other hand, the static intrinsic spin currents are present
in 2D hole systems described by the effective ${\bf p}$-cubic SO coupling
Hamiltonian,$^{11}$ as demonstrated by theoretical studies.$^{12-16}$
It is believed that the experimentally observed spin Hall effect for 2D
holes$^{4,5}$ is of the intrinsic origin.

In spite of the fact that experiments on spin excitation are often carried
out in the presence of magnetic fields, the theoretical work devoted to the
influence of a magnetic field on the electric-field-induced spin currents is
very limited.$^{17}$ For 2D hole systems, where the intrinsic spin
currents exist, the spin-Hall conductivity has been calculated$^{18,19}$
using Kubo formalism in the collisionless approximation. Although this approach
allows one to study the case of strong magnetic field and to describe
Shubnikov-de Haas oscillations of spin conductivity,$^{19}$ it cannot be used
for weak magnetic fields, when the cyclotron frequency is comparable to or
less than the momentum relaxation rate due to scattering. In particular, the
zero magnetic field limit of the spin-Hall conductivity appears to be singular
in the collisionless approximation.$^{18}$ To describe the region of weak
magnetic fields, which is the most important for experimental studies,
a consideration of spin conductivity in the presence of scattering is necessary.

In this Brief Report, the intrinsic spin current is calculated for the case of a
classically weak magnetic field ${\bf H}$ perpendicular to the 2D layer. The
elastic scattering of carriers is taken into account. The general approach to
the problem assuming arbitrary SO coupling Hamiltonian is followed by
application to 2D hole systems described by the ${\bf p}$-cubic SO coupling.
The analytical solution obtained below shows that the spin-Hall conductivity increases
at small $H$ and has a maximum in the region where the cyclotron frequency is
smaller than the relaxation rate. Moreover, it is found that the diagonal
component of the spin conductivity appears. For this reason, the spin response
in Hall bars should be described by the effective spin-Hall conductivity which
is a combination of diagonal and non-diagonal components of the spin conductivity tensor.

Consider 2D quasiparticles whose states are doubly degenerate in spin if the magnetic
field and SO interaction are absent. In the presence of both the magnetic field and
SO interaction, the single-particle Hamiltonian is
%1
\begin{equation}
\hat{H} = \frac{\bpi^2}{2m} + \hat{h}_{\sbpi} + V_{\bf r},~~~ \hat{h}_{\sbpi}=\hbar
\bOmega_{\sbpi} \cdot \hat{\bsigma},
\end{equation}
where $\bpi=-i\hbar \nabla_{\bf r}-(e/c){\bf A}_{\bf r}$ is the operator of
momentum in the 2D plane ($xy$) and ${\bf A}_{\bf r}$ is the vector potential describing
the magnetic field according to ${\bf H}=[\nabla_{\bf r} \times {\bf A}_{\bf r}]$.
Next, $m$ and $e$ are the effective mass and electric charge of the quasiparticles,
$c$ is the velocity of light, $V_{\bf r}$ is the scattering potential, and
$\hat{\bsigma}$ is the vector of Pauli matrices. The $2 \times 2$ matrix term
$\hat{h}_{\sbpi}$ describes both intrinsic SO coupling and Zeeman interaction.
The extrinsic SO coupling effects are not considered.

The limit of weak (classical) magnetic field corresponds to the condition
$\hbar \omega_c \ll \overline{\varepsilon}$, where $\omega_c=|e|H/mc$ is the
cyclotron frequency and $\overline{\varepsilon}$ is the mean kinetic energy of
quasiparticles. It is convenient to describe transport phenomena by using
the kinetic equation for the Wigner distribution function $\hat{\rho}_{{\bf pr}t}$,
which is a $2 \times 2$ matrix over the spin indices (see, for example, Refs. 20
and 21) and depends on the 2D momentum ${\bf p}$, coordinate ${\bf r}$, and
time $t$. For the spatially-homogeneous and static case considered below, the
coordinate and time dependence is omitted. Searching for the linear response
to the applied electric field ${\bf E}$, one represents the distribution
function in the form $\hat{\rho}_{\mathbf{p}}=\hat{f}^{(eq)}_{\mathbf{p}}+
\hat{f}_{\mathbf{p}}$, where $\hat{f}_{\bf p}$ is the non-equilibrium part
satisfying the linearized kinetic equation$^{20,21}$
%2
\begin{equation}
\frac{i}{\hbar} \left[ \hat{h}_{\bf p},\hat{f}_{\bf p} \right]
+\frac{e}{c}\left\{[\hat{{\bf u}}({\bf p}) \times {\bf H}],
\frac{\partial \hat{f}_{\bf p}}{\partial {\bf p}} \right\}
+ e {\bf E} \cdot \frac{\partial \hat{f}^{(eq)}_{\mathbf{p}}}{\partial
{\bf p}} = \widehat{J}( \hat{f}| \mathbf{p} ),
\end{equation}
$\hat{{\bf u}}({\bf p})= {\bf v}_{\bf p}+\partial{\hat{h}_{\bf p}}/\partial
{\bf p}$ is the group velocity in the presence of spin-orbit interaction, ${\bf
v}_{\bf p}={\bf p}/m$, and $\{\hat{a},\hat{b}\}= (\hat{a} \hat{b}+\hat{b}\hat{a})/2$
denotes the symmetrized matrix product. The collision integral $\widehat{J}$
describing the elastic scattering is considered under the assumption
that the relaxation rate is small in comparison to
$\overline{\varepsilon}/\hbar$. Assuming that the spin-splitting energy
$2 \hbar |\bOmega_{\bf p}|$ is small in comparison to $2 \overline{\varepsilon}$,
it is convenient to expand the collision integral in series of
$\hbar |\bOmega_{\bf p}|/\overline{\varepsilon}$. Retaining the
first-order terms in this expansion, one obtains$^{21,22}$
%3
\begin{eqnarray}
\widehat{J} ( \hat{f}| \mathbf{p} ) \simeq
\frac{2 \pi}{\hbar} \int \! \! \frac{d {\bf p'}}{(2 \pi \hbar)^2}
w_{|{\bf p}-{\bf p}'|} \left[ (\hat{f}_{\bf p'}- \hat{f}_{\bf p})
\delta (\varepsilon_{p'} -\varepsilon_{p}) \right. \nonumber \\
\left. - \{ (\hat{h}_{\bf p}-\hat{h}_{\bf p'}),
(\hat{f}_{\bf p'}- \hat{f}_{\bf p}) \}
\frac{\partial \delta (\varepsilon_{p'} -\varepsilon_{p}) }{\partial
\varepsilon_{p'}} \right],
\end{eqnarray}
where $w_{|{\bf p}-{\bf p}'|}$ is the spatial Fourier transform of the correlation
function of the scattering potential $V_{\bf r}$, and $\varepsilon_{p}=p^2/2m$ is the
kinetic energy. The condition $\hbar |\bOmega_{\bf p}| \ll  \overline{\varepsilon}$ also
allows one to write the equilibrium distribution function in the form $\hat{f}^{(
eq)}_{\mathbf{p}} \simeq f_{\varepsilon_p} + \hat{h}_{\bf p} f'_{\varepsilon_p}$,
where $f'_{\varepsilon_p} \equiv \partial f_{\varepsilon_p}/ \partial
\varepsilon_p$ and $f_{\varepsilon}$ is the Fermi distribution function.

The calculations below are done in the approximation of short-range
scattering potential, when $w_{|{\bf p}-{\bf p}'|} \simeq w$ is constant.
It is convenient to use the spin-vector representation $\hat{f}_{\bf p}=
{\rm f}^0_{\bf p}+ \hat{\bsigma} \cdot {\rm {\bf f}}_{\bf p}$.
In the leading order with respect to the small parameter $\hbar |\bOmega_{\bf
p}|/\varepsilon_p$ one finds
%4
\begin{equation}
{\rm f}^0_{\bf p} \simeq -G_{\bf p} f'_{\varepsilon_p},~~~ G_{\bf p}=
e \frac{\nu ({\bf E} \cdot {\bf v}_{\bf p})
- \omega_c [{\bf E} \times {\bf v}_{\bf p}] \cdot {\bf n} }{\nu^2+\omega_c^2},
\end{equation}
where $\nu=mw/\hbar^3$ is the relaxation rate and ${\bf n}=(0,0,1)$ is the
unit vector in the direction of ${\bf H}$.
Here and below, $e$ is taken to be positive, since the theory will be applied to
hole systems. With the use of Eq. (4), the equation for the vector-function
${\rm {\bf f}}_{\bf p}$ is written as
%5
\begin{equation}
-2 [ \bOmega_{\bf p} \times {\rm {\bf f}}_{\bf p}] - \omega_c \frac{\partial}{
\partial \varphi} {\rm {\bf f}}_{\bf p} + {\bf R}_{\bf p} =
\nu (\overline{ {\rm {\bf f}}}_{\bf p}-{\rm {\bf f}}_{\bf p}),
\end{equation}
where $\varphi$ is the angle of the vector ${\bf p}$, and the line over a function
denotes the angular average $(2 \pi)^{-1} \int_0^{2 \pi} d \varphi \ldots~$.
The components of the vector-function ${\bf R}_{\bf p}$ are
%6
\begin{eqnarray}
R^{\alpha}_{\bf p}= \hbar \nu \frac{\partial}{\partial \varepsilon_p}
[(\Omega^{\alpha}_{\bf p} G_{\bf p} -\overline{\Omega^{\alpha}_{\bf p} G_{\bf p}}
)f'_{\varepsilon_p} ] ~~~~~~\nonumber \\
+\hbar e [{\bf E} \times {\bf n}]
\cdot \frac{\partial {\bf p} \Omega^{\alpha}_{\bf p}
}{\partial \varphi} \frac{f'_{\varepsilon_p}}{p^2} -
\hbar \omega_c \frac{\partial}{\partial \varphi} \left[
\Omega^{\alpha}_{\bf p} \frac{\partial G_{\bf p}
f'_{\varepsilon_p}}{\partial \varepsilon_p} \right].
\end{eqnarray}

Equations (5) and (6) are valid for arbitrary $\bOmega_{\bf p}$. The first and
the second terms in Eq. (5) describe spin precession and cyclotron motion,
respectively. The term ${\bf R}_{\bf p}$ describes excitation of spins
by the electric field. Solution of Eq. (5) determines the non-equilibrium
spin current density conventionally defined as
%7
\begin{equation}
{\bf q}_{\gamma}= S \int \frac{d {\bf p}}{(2 \pi
\hbar)^2} {\rm Tr}( \{ \hat{\bsigma} , \hat{u}_{\gamma}({\bf p}) \}
\hat{f}_{\bf p} ),
\end{equation}
where $S=1/2$ for electrons and $S=3/2$ for holes in the ground-state subband.
Applying Eqs. (5)-(7) to electron system described by a ${\bf p}$-linear SO
coupling Hamiltonian, one can find that the spin current is zero in the absence of
Zeeman coupling, as expected.$^{9}$ Let us consider the 2D holes near the bottom
of the ground-state subband in an asymmetric [001]-grown quantum well. By adding
the Zeeman term $-S \omega_{ H} \hat{\sigma}_z$ to the effective ${\bf p}$-cubic
SO coupling Hamiltonian$^{11}$ describing the 2D holes, one gets
\begin{equation}
\bOmega_{\bf p} =(\kappa p^3 \cos 3 \varphi,~ \!
\kappa p^3 \sin 3 \varphi,~ \! -3\omega_{ H}/2),
\end{equation}
where $\kappa$ is a constant determined by the Luttinger parameters
and confinement potential. Next, $\omega_{ H}=g \mu_{ B}
H/\hbar$, where $g$ is the effective g-factor of holes, and $\mu_{ B}=|e|
\hbar/2 m_0 c$ is the Bohr magneton. Expanding ${\rm {\bf f}}_{\bf p}$ in
series of angular harmonics, one can solve Eq. (5) exactly and obtain the result
\begin{equation}
{\bf q}_{\gamma}=(0,0,q_{\gamma}),~~~q_{\gamma}=\Sigma_{\gamma \beta} E_{\beta}.
\end{equation}
The tensor of spin conductivity, $\Sigma_{\gamma \beta}$, contains two contributions:
\begin{equation}
\Sigma_{\gamma \beta}=\Sigma^{ (0)}_{\gamma \beta}+\Sigma^{ (1)}_{\gamma \beta},
\end{equation}
\begin{equation}
\Sigma^{ (0)}_{\gamma \beta}= \frac{9 e\omega_{ H}}{4 \pi \hbar} ~ \!
\frac{\nu \delta_{\gamma \beta}+\omega_c e_{z \gamma \beta}}{\nu^2+\omega_c^2}
\int_0^{\infty} \! \! d \varepsilon_p (-f'_{\varepsilon_p}),
\end{equation}
\begin{eqnarray}
\Sigma^{ (1)}_{\gamma \beta}=\frac{9e}{8 \pi \hbar}
\int_0^{\infty} \!\!\! d \varepsilon_p (-f'_{\varepsilon_p}) {\rm Re} \frac{[i \delta_{\gamma\beta}
+ e_{z \gamma \beta} ]\Delta_p^2 \nu \nu_-}{(\nu \! - \! i\omega_c)^2 (\nu_+ \nu_- \! + \! \Delta_p^2)},\\
\nu_{\pm}=\nu - i\omega_c \pm 3 i (\omega_c-\omega_{ H}).~~~~~~~~~~~~ \nonumber
\end{eqnarray}
In these equations, $e_{z \gamma \beta}$ is the antisymmetric unit tensor
and $\hbar \Delta_p=2 \hbar \kappa p^3$ is the SO splitting energy.
Notice the symmetry properties $\Sigma_{xy}=-\Sigma_{yx}$ and $\Sigma_{xx}=\Sigma_{yy}$.
The contribution $\Sigma^{ (0)}_{\gamma \beta}$ exists owing to equilibrium
spin polarization by the magnetic field in the presence of Zeeman coupling. Indeed,
the equilibrium spin density is given by $s_z=S \int \frac{d {\bf p}}{(2 \pi \hbar)^2}
{\rm Tr}(\hat{\sigma}_z \hat{f}^{(eq)}_{\bf p} ) \simeq 9 \omega_{ H} m/4
\pi \hbar$, and the related spin conductivity is $\Sigma^{ (0)}_{\gamma \beta}=
(s_z/e n_{h}) \sigma_{\gamma \beta}$, where $\sigma_{\gamma \beta}$ is the Drude
conductivity tensor and $n_{h}$ is the hole density. In contrast, the contribution
$\Sigma^{ (1)}_{\gamma \beta}$ is caused by the SO coupling. In zero magnetic field,
when $\omega_c = \omega_{ H} =0$, $\Sigma^{ (0)}_{\gamma \beta}$ disappears
and $\Sigma^{ (1)}_{\gamma \beta}$ contains only non-diagonal (Hall) component,
$\Sigma^{ (1)}_{xy}=9e \Gamma/8 \pi
\hbar$ (see Refs. 12-16), where the factor $\Gamma$ describes suppression of
the spin-Hall conductivity by the disorder. In the limit of low
temperature, when the hole gas is degenerate, $\Gamma=[1+ (\nu/\Delta)^2]^{-1}$,
where $\hbar \Delta \equiv \hbar \Delta_{p_F}$ is the SO splitting energy at the Fermi surface
and $p_{ F}=\sqrt{2m \varepsilon_{ F}}$ is the Fermi momentum. The transition
to the low-temperature limit in Eqs. (11) and (12) implies $-f'_{\varepsilon_p}=
\delta(\varepsilon_p-\varepsilon_{ F})$, so the integral in Eq. (11) is equal
to unity and the integration over energy in Eq. (12) is reduced to the substitution
$\Delta_p \rightarrow \Delta$.

\begin{figure}[ht]
\begin{center}
\includegraphics[scale=0.4]{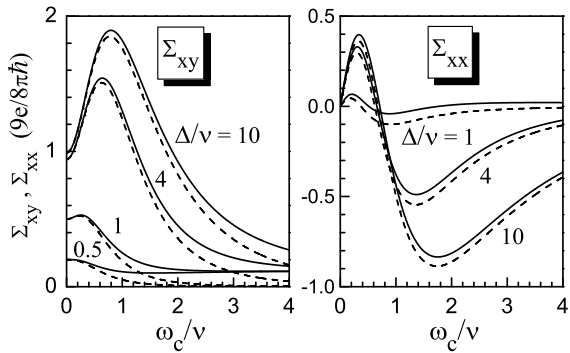}
\end{center}
\addvspace{-0.7 cm}\caption{The magnetic-field dependence of the spin
conductivity components $\Sigma_{xy}$ and $\Sigma_{xx}$ (solid) and of their
SO parts $\Sigma_{xy}^{ (1)}$ and $\Sigma_{xx}^{ (1)}$ (dash) for degenerate
hole gas.}
\end{figure}

The main features of the spin conductivity in the magnetic field are the presence
of both non-diagonal and diagonal components and the unusual (non-Drude) dependence
of these components on the ratio $\omega_c/\nu$. Figure 1 shows the plots
of $\Sigma_{xy}$ and $\Sigma_{xx}$ as functions of this ratio for several values
of $\Delta/\nu$. Notice that the plots for $\Delta/\nu=4$ correspond to experimental
conditions of Ref. 5. The ratio $\omega_{ H}/\omega_c=g m/2m_0$ is estimated as
0.06, using $g =0.44$ and $m/m_0 \simeq 0.27$ given in Ref. 19. Since this ratio
is small, the influence of Zeeman coupling on $\Sigma_{xy}^{ (1)}$ and $\Sigma_{xx
}^{ (1)}$ is weak. However, $\Sigma^{ (1)}_{xy}$ is considerably suppressed with
the increase of $\omega_c$, while $\Sigma^{ (0)}_{xy}$ saturates at the value
$(9e/8 \pi \hbar) (g m/m_0)$ determined by the g-factor and effective mass.
Therefore, the contribution $\Sigma^{ (0)}_{xy}$ dominates in $\Sigma_{xy}$ at
larger $\omega_c/\nu$, especially for the dirty case $\Delta < \nu$, when the contribution
$\Sigma^{ (1)}_{xy}$ is suppressed. It is remarkable that $\Sigma_{xy}$
is a non-monotonic function of the magnetic field and increases in the region of weak
fields. This behavior takes place in the clean regime $\Delta \gg \nu$, when the spin
splitting is not suppressed by the disorder. In the limit $\Delta \gg \nu,
3 \omega_c$, the contribution $\Sigma^{ (1)}_{xy}$ is independent of the spin
splitting:
\begin{equation}
\Sigma^{ (1)}_{xy} \simeq \frac{9e}{8 \pi \hbar} \frac{\nu^2}{\nu^2+ \omega^2_c}
\left[1+ \frac{6 \omega_c (\omega_c-\omega_{ H})}{\nu^2+ \omega^2_c} \right].
\end{equation}
According to this expression, the maximum $\Sigma^{ (1)}_{xy}$ is approximately
twice larger than the zero-field spin-Hall conductivity $9e/8 \pi \hbar$. As
follows from the Onsager symmetry principle, $\Sigma_{xy}$ is symmetric with
respect to the magnetic field reversal. In contrast, the diagonal component
$\Sigma_{xx}$ is antisymmetric in ${\bf H}$. This component is absent at
$H=0$ for the particular case of ${\bf p}$-cubic SO coupling considered
here. As seen from Fig. 1, $\Sigma_{xx}$ changes its sign at a finite magnetic
field and is strongly suppressed by the disorder, though the magnetic-field
suppression of this component is weak. The behavior of $\Sigma^{ (1)}_{xx}$ in
the limit $\Delta \gg \nu, 3 \omega_c$ is described by the expression
\begin{equation}
\Sigma^{ (1)}_{xx} \simeq \frac{9e}{8 \pi \hbar} \frac{\nu \omega_c}{\nu^2+ \omega^2_c}
\left[2-3 \frac{\omega_{ H}}{\omega_c} -
\frac{6 \omega_c (\omega_c-\omega_{ H})}{\nu^2+ \omega^2_c} \right].
\end{equation}

\begin{figure}[ht]
\begin{center}
\includegraphics[scale=0.36]{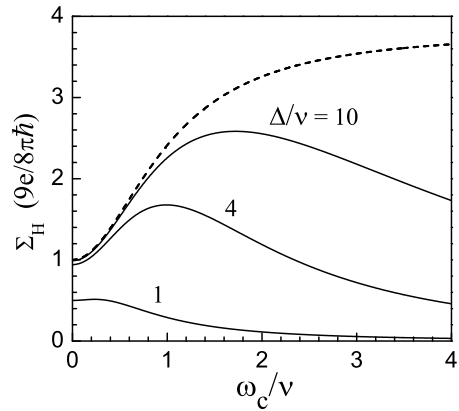}
\end{center}
\addvspace{-0.7 cm}\caption{The magnetic-field dependence of the effective
spin-Hall conductivity $\Sigma_{H}$ for the Hall bar geometry. The dashed
line corresponds to the limit $\Delta \gg \nu, 3 \omega_c$ given by Eq. (16).}
\end{figure}

In the experiments using the Hall bar geometry, the electric field acting
on the carriers has both longitudinal and transverse (Hall) components,
the latter is determined by the requirement of zero electric current in the
transverse direction. If the field $E_y$ is applied along the bar, the Hall
field is $E_x=-(\sigma_{xy}/\sigma_{xx})E_y=-(\omega_c/\nu)E_y$, and the
spin-Hall current $q_x=\Sigma_H E_y$ is described by the {\em effective}
spin-Hall conductivity
\begin{equation}
\Sigma_H \! = \! \Sigma_{xy} \! - \! \frac{\omega_c}{\nu}\Sigma_{xx} \! = \!
\frac{9e}{8 \pi \hbar} {\rm Re} \frac{\Delta^2 \nu_-}{(\nu \! - \! i\omega_c)
(\nu_+ \nu_- \! + \! \Delta^2)}
\end{equation}
written here for the case of degenerate hole gas. The Zeeman part
$\Sigma^{ (0)}_{\gamma \beta}$ does not contribute to $\Sigma_H$. The electric
current through the Hall bar is $j_y=\sigma_0 E_y$, where $\sigma_0=e^2n_h/m\nu$
is the Drude conductivity at $H=0$. Therefore, the quantity $\Sigma_H/\sigma_0$
determines the ratio of spin-Hall current to electric current in Hall bars.
Figure 2 shows a considerable increase of $\Sigma_{H}$ with respect to its
zero-field value and a non-monotonic behavior in the clean regime. If $\Delta
\gg \nu, 3 \omega_c$, one finds a simple expression
\begin{equation}
\Sigma_H \simeq \frac{9e}{8 \pi \hbar} \left[1+
\frac{3 \omega_c (\omega_c-\omega_{ H})}{\nu^2+ \omega^2_c} \right].
\end{equation}
The Hall bar also carries the longitudinal spin current $q_y=\Sigma_D E_y$
expressed through the effective diagonal spin conductivity $\Sigma_D = \Sigma_{xx}
+ (\omega_c/\nu)\Sigma_{xy}$.

Since the approach used in this Brief Report does not describe the Shubnikov-de Haas
oscillations, the results presented above can be directly applied if these
oscillations are suppressed either by the temperature or by the disorder.
In the general case, Eqs. (10)-(16) should be treated as the expressions for
the slow envelope part of the oscillating spin conductivity. If the relaxation
rate $\nu$ is aimed to zero, $\Sigma^{ (1)}_{xy}$ disappears at a finite
magnetic field, in agreement with the result of Ref. 18. The effective
spin-Hall conductivity $\Sigma_{H}$, however, remains finite in this limit.

To conclude, in the presence of a perpendicular magnetic field the spin current
of 2D holes is described by the spin conductivity tensor containing both diagonal
($\Sigma_{xx}$) and non-diagonal ($\Sigma_{xy}$) components. These components are
non-monotonic functions of $H$ and increase in the low-field region. The enhancement of
the spin-Hall component $\Sigma_{xy}$ in comparison to its zero-field value and the
appearance of the diagonal component $\Sigma_{xx}$ are explained by the asymmtery
introduced by the Lorentz force, when the Fermi surface is shifted in the direction
determined by the angle $\arctan(\omega_c/\nu)$ with respect to ${\bf E}$; see
Eq. (4). As the magnetic field increases and $\omega_c$
exceeds $\nu$, the components $\Sigma_{xx}$ and $\Sigma_{xy}$ are suppressed
by the field. The consideration of the component $\Sigma_{xy}$ alone is not
sufficient for description of the spin-Hall response in a Hall bar. It is necessary
to introduce the effective spin-Hall conductivity $\Sigma_H$ which determines
the transverse spin current $q_x=\Sigma_H E_y$ proportional to the applied longitudinal
electric field $E_y$. Since the spin density accumulation near the edges of the Hall
bar is estimated as$^5$ $s^{edge}_z \sim q_x m/p_{ F}$, the behavior of
$\Sigma_H$ can be directly investigated by measuring the magnetic-field
dependence of this accumulation. The theory suggests (see Fig. 2)
that $\Sigma_H$ is considerably enhanced by the magnetic field in the clean
systems, where the spin coherence is not suppressed by the disorder. This
condition is attainable in the existing samples$^{4,5}$ and can be improved
by increasing the hole density, because $\Delta \propto n_h^{3/2}$. If
$\Delta/\nu \simeq 4$ (see Ref. 5), $\Sigma_H$ has a maximum at $\omega_c \simeq
\nu$, which corresponds to $H \simeq 2.8$ T if one uses the parameters
$m=0.27~m_0$ and $\hbar \nu = 1.2$ meV typical for 2D holes in GaAs quantum
wells.$^{4,5}$ Therefore, experimental verification of the theoretical results
is possible and desirable.

\end{document}